\documentclass[3p,12pt]{elsarticle}

\usepackage{lineno}
\modulolinenumbers[5]

\usepackage{scalerel}

\usepackage{amsfonts,amsmath,amssymb,graphicx}
\usepackage{color,mathrsfs,bm,appendix,subfig}
\usepackage[dvipsnames]{xcolor}

\usepackage[figuresright]{rotating}
\usepackage{multirow}
\usepackage{colortbl}

\usepackage[colorlinks=true]{hyperref}

\usepackage{epstopdf}
\usepackage{capt-of}

\usepackage{booktabs}

\begin{document}

\begin{frontmatter}

\title{High-Order Implicit Hybridizable Discontinuous Galerkin Method for the Boltzmann Equation}

\author{Wei Su, Peng Wang, Yonghao Zhang, Lei Wu\corref{mycorrespondingauthor}}
\address{James Weir Fluids Laboratory, Department of Mechanical and Aerospace Engineering, \\University of Strathclyde, G1 1XJ Glasgow, United Kingdom}
\cortext[mycorrespondingauthor]{Corresponding author: lei.wu.100@strath.ac.uk}




\begin{abstract}
The high-order hybridizable discontinuous Galerkin (HDG) method combining with an implicit iterative scheme is used to find the steady-state solution of the Boltzmann equation with full collision integral on two-dimensional triangular meshes. The velocity distribution function and its trace are approximated in the piecewise polynomial space of degree up to 4. The fast spectral method (FSM) is incorporated into the DG discretization to evaluate the collision operator. Specific polynomial approximation is proposed for the collision term to reduce the computational cost. The proposed scheme is proved to be accurate and efficient.

\end{abstract}


\end{frontmatter}


\section{Introduction}
In gas kinetic theory, the motion of molecules in dilute gas is mathematically described by one-particle velocity distribution function (VDF) and the macroscopic flow properties are derived from the velocity moment of the VDF. In Boltzmann's description, all molecules modeled by a nonlinear collision operator that is a fivefold integral with three dimensions in velocity space and tow dimensions in a unit sphere. The multi-dimensional structure of the collision operator posed a real challenge to the numerical solution of the Boltzmann equation.

The high-order Runge-Kutta discontinuous Galerkin (RKDG) method has been applied to solve the Boltzmann kinetic model equations, where the collision integral is reduced to simpler relaxation terms~\cite{SU2015123}. Numerical tests have shown that, although the second-order RKDG method is faster than a second-order Runge-Kutta finite volume method (FVM) by one order of magnitude, the third-order RKDG scheme is not more efficient. This is mainly due to two facts: 1) higher-order method involves larger number of degrees of freedom (DoF); 2) the iterative time interval which is restricted by the Courant-Friedrichs-Lewy (CFL) condition is smaller in higher-order method. Implicit scheme could be employed to relax the CFL restriction. However, classical DG methods are computationally expensive for steady or implicit solvers, since the number of globally DoF is significantly high~\cite{Moghtader2016}.

In recent years, a new DG method called Hybridizable discontinuous Galerkin (HDG) method has been developed with the aim to reduce the number of DoF~\cite{Cockburn2009a}. By producing a final system in terms of the degrees of freedom in approximating traces of the field variables, HDG could significantly reduce the number of global unknowns, since the traces are defined on cell interfaces and single-valued. This advantage is prominent for the gas kinetic simulation, where a cumbersome system of governing equations need to be solved. The authors has applied the HDG method for the solution of kinetic model equation~\cite{Su2018}. Compared to the classical DG method, it is found that the number of DoF in HDG is smaller when the degree of approximation polynomials is larger than 1, and is more efficient. The higher order and more triangles, the more significant this difference will be. Actually, the number of DoF of the HDG becomes closer to that of the continuous finite element method for higher orders~\cite{Huerta2013}.

In this work, we extend the HDG formulation to the Boltzmann equation with full collision operator. The remainder of the paper is organized as follows. In Sec.~\ref{Boltzmann}, the Boltzmann equation and the fast spectral method (FSM) that is used to evaluate the collision operator are introduced. In Sec.~\ref{HDG}, the HDG method is described with details in the formulation of the collision operator. Two different problems are simulated in Sec.~\ref{Results} to assess the accuracy and efficiency of the proposed scheme. Conclusions are presented in Sec.~\ref{Concludsion}.

\section{The Boltzmann equation}~\label{Boltzmann}

In gas kinetic theory, variation of VDF $f\left(t,\bm x,\bm v\right)$ in dependence of the time $t$, the spatial position $\bm x\left(x_1,x_2,x_3\right)$ and the molecular velocity $\bm v\left(v_1,v_2,v_3\right)$ is governed by the Boltzmann equation. Neglecting external force, the Boltzmann equation for a single-species monatomic gas is written in the following dimensionless form:
\begin{equation}
\frac{\partial f}{\partial t} + \bm v\cdot\frac{\partial f}{\partial\bm x}=\mathcal{C}\left(f,f_{*}\right),
\end{equation}
where, VDF is defined as that the quantity $f\left(t,\bm x,\bm v\right)\mathrm{d}\bm x\mathrm{d}\bm v$ is the number of molecules in the phase-space volume $\mathrm{d}\bm x\mathrm{d}\bm v$. $\mathcal{C}\left(f,f_{*}\right)$ is the collision operator, which can be split into the gain term $\mathcal{C}_{+}$ and loss term $\mathcal{C}_{-}$ as:
\begin{equation}
\mathcal{C}\left(f,f_{*}\right)=\mathcal{C}_{+}-\mathcal{C}_{-}
=\int\int B\left(\theta,|\bm v-\bm v_{*}|\right)f\left(\bm v'_{*}\right)f\left(\bm v'\right)\mathrm{d}\Omega\mathrm{d}\bm v_{*}-\nu f.
\end{equation}
where
\begin{equation}
\nu=\int\int B\left(\theta,|\bm v-\bm v_{*}|\right)f\left(\bm v_{*}\right)\mathrm{d}\Omega\mathrm{d}\bm v_{*},
\end{equation}
is the collision frequency. Here, $B\left(\theta,|\bm v-\bm v_{*}|\right)$ is the collision kernel; $\bm v$, $\bm v_{*}$ are the pre-collision molecular velocities of a collision pair, and $\bm v'$, $\bm v'_{*}$ are the corresponding post-collision molecular velocities; $\Omega$ is the unit vector along the relative post-collision velocity $\bm v'-\bm v'_{*}$; $\theta$ is the deflection angle between the pre- and post-collision relative velocities. For simplicity the time and spatial position is omitted in writing the collision operator.

All the macroscopic quantities, such as the number density $n$, bulk velocity $\bm u\left(u_1,u_2,u_3\right)$, temperature $T$, pressure tension $\bm P$ and heat flux $\bm Q\left(Q_1,Q_2,Q_3\right)$ can then be calculated via the velocity moments of the distribution function:
\begin{equation}
\begin{aligned}
n = \int f\mathrm{d}\bm v,\quad\bm u=\frac{1}{n}\int\bm vf\mathrm{d}\bm v,\quad T=\frac{2}{3n}\int|\bm v-\bm u|^2f\mathrm{d}\bm v,\\
\bm P=2\int\left(\bm v-\bm u\right)\otimes\left(\bm v-\bm u\right)f\mathrm{d}\bm v,\quad \bm Q=\int\left(\bm v-\bm u\right)|\bm v-\bm u|^2f\mathrm{d}\bm v.
\end{aligned}
\label{Macro}
\end{equation}

The above dimensionless variables are non-dimensionalized as: $\bm x$ is normalized by a characteristic flow length $H$; $T$ is normalized by a reference temperature $T_0$; $n$ is normalized by the average number density $n_0$ at $T_0$; $\bm v$ and $\bm u$ are normalized by the most probable speed $v_\text{m}=\sqrt{2k_\text{B}T_0/m}$ with $k_{\text{B}}$ and $m$ being the Boltzmann constant and molecular mass; $t$ is normalized by $H/v_\text{m}$; $f$ is normalized by $n_0/v^3_\text{m}$; $\bm P$ is normalized by $n_0k_\text{B}T_0$; and $q_i$ is normalized by $n_0k_\text{B}T_0v_\text{m}$.

The collision kernel $B\left(\theta,|\bm v-\bm v_{*}|\right)$ is always non-negative and depends on the modules of the pre-collision relative velocity and the deflection angle. The form of $B$ is only determined when a certain intermolecular potential is given~\cite{Chapman1970}. One of the most widely used phenomenological models is the inverse power law (IPL) potential, however, its total collision cross-section is infinite at the grazing collision limit, i.e. $\theta\rightarrow0$. In practice, simplified collision kernel is adopted with the aim to eliminate the infinity and recover the correct transport coefficients. Commonly used ones are the well-known variable hard sphere (VHS) model~\cite{Bird1994} and variable soft sphere (SSH) model~\cite{Koura1991}. In this paper, the collision kernel is modeled as~\cite{WU2013,wu2014}:
\begin{equation}
B\left(\theta,|\bm v-\bm v_{*}|\right)=\frac{5|\bm v-\bm v_{*}|^{2\left(1-\omega\right)}}{2^{7-\omega}\Gamma\left(\frac{5-2\omega+\gamma}{2}\right)\Gamma\left(2-\frac{\gamma}{2}\right)Kn}\sin^{1-2\omega+\gamma}\left(\frac{\theta}{2}\right)\cos^{-\gamma}\left(\frac{\theta}{2}\right),
\end{equation}
where, $\Gamma$ is the Gamma function, $\gamma$ is a free parameter, $\omega$ is the viscosity index (i.e. the shear viscosity $\mu$ of the gas is proportional to $T^{\omega}$) and $Kn$ is the unconfined Knudsen number given at the reference condition:
\begin{equation}
Kn=\frac{\mu\left(T=T_0\right)}{n_0H}\sqrt{\frac{\pi}{2mk_{\text{B}}T_0}}.
\end{equation}
This specific type of collision kernel could describe all IPL potentials (except the Coulomb potential) and recover not only the value of the shear viscosity but also the correct ratio between the coefficients of shear stress and diffusion. It is worthy mentioning that other intermolecular potentials, such as the Lennard-Jones potential, Coulomb potential and rigid attract potential could be easily incorporated~\cite{WU2013,wu2014,Wu2015}.

\subsection{The fast spectrum method}
The collision operator is a fivefold integral with three dimensions in the molecular velocity space and two dimensions in a unit sphere. In this paper, we apply the fast spectrum method to evaluate the collision operator in the frequency space. The VDF is periodized on a truncated domain $\mathcal{D}=[-L,L]^3$ and expanded in Fourier series with $N=N_1\times N_2\times N_3$ components:
\begin{equation}
f\left(t,\bm x,\bm v\right)=\sum^{\bm N/2-1}_{\bm j=-\bm N/2}\bar{f}^{\bm j}\left(t,\bm x\right)\exp\left(\imath\bm\xi^{\bm j}\cdot\bm v\right),
\end{equation}
\begin{equation}
\bar{f}^{\bm j}\left(t,\bm x\right)=\frac{1}{\left(2L\right)^3}\int_{\mathcal{D}}f\left(t,\bm x,\bm v\right)\exp\left(-\imath\bm\xi^{\bm j}\cdot\bm v\right)\mathrm{d}\bm v,
\label{Spectrum}
\end{equation}
where $\imath$ is the imaginary unit, $\bm N=\left(N_1,N_2,N_3\right)$, $\bm\xi^{\bm j}=\bm j\pi/L$ with $\bm j=\left(j_1,j_2,j_3\right)$ is the discrete frequencies, $\bar{f}^{\bm j}$ is the spectrum of the VDF and $L$ is the maximum truncated velocity. In order to take advantage of FFT, the discretization in frequency necessitate being uniformly distributed.

The gain term in collision integral and the collision frequency are evaluated through expanding in Fourier series:
\begin{equation}
\mathcal{C}_{+}=\sum_{\bm j=-\bm N/2}^{\bm N/2-1}\bar{\mathcal{C}}^{\bm j}_{+}\exp\left(\imath\bm\xi^{\bm j}\cdot\bm v\right),
\end{equation}
\begin{equation}
\nu=\sum_{\bm j=-\bm N/2}^{\bm N/2-1}\bar{\nu}^{\bm j}\exp\left(\imath\bm\xi^{\bm j}\cdot\bm v\right),
\end{equation}
where the $\bm j$-th Fourier modes are related to the VDF spectrum as follows~\cite{WU2013,wu2014}:
\begin{equation}
\bar{\mathcal{C}}_{+}^{\bm j}=\sum^{\bm N/2-1}_{\substack{\bm l+\bm m=\bm j\\\bm l,\bm m=-\bm N/2}}\bar{f}^{\bm l}\bar{f}^{\bm m}\beta\left(\bm l,\bm m\right),\quad \bar{\nu}^{\bm j}=\bar{f}^{\bm j}\beta\left(\bm j,\bm j\right),
\end{equation}
where, $\beta$ is the collision kernel mode, which is related to the integrals in a sphere supporting the VDF. Its $\left(\bm l,\bm m\right)$-th component is approximated through $M_\text{qua}$-point Gauss-Legendre quadrature as:
\begin{equation}
\begin{aligned}
\beta\left(\bm l,\bm m\right)\simeq\frac{20}{2^{7-\omega}\Gamma\left(\frac{5-2\omega+\gamma}{2}\right)\Gamma\left(2-\frac{\gamma}{2}\right)Kn}\cdot\\
\sum^{M_\text{qua}}_{p,q=1}\sin\left(\theta_p\right)\Psi\left(\sqrt{|\bm\xi^{\bm m}|^2-\left(\bm\xi^{\bm m}\cdot\bm e_{p,q}\right)^2}\right)\Phi\left(\bm\xi^{\bm l}\cdot\bm e_{p,q}\right)\varpi_p\varpi_q,
\end{aligned}
\end{equation}
where $\bm e_{p,q}=\left(\sin\theta_p\cos\phi_q,\sin\theta_p\sin\phi_q,\cos\theta_p\right)$; $\theta_p$ ($\phi_q$) and $\varpi_p$ ($\varpi_p$) are the $p$ ($q$)-th point and weight in the Gauss-Legendre quadrature, respectively, with $\theta$, $\phi\in\left[0,\pi\right]$. The functions $\Psi$ and $\Phi$ are define as:
\begin{equation}
\Psi\left(a\right)=2\pi\int_0^R\rho^{1-\gamma}J_0\left(\rho a\right)\mathrm{d}\rho,\quad\Phi\left(a\right)=2\int_0^R\rho^{2\left(1-\omega\right)+\gamma}\cos\left(\rho a\right)\mathrm{d}\rho,
\end{equation}
where $J_0$ is the zeroth-order Bessel function, and $R$ is the radius of the sphere to support the VDF, which is chosen approximately as $R=2\sqrt{2}L/(2+\sqrt{2})$~\cite{WU2013}.

Note that integral with respect to the velocity space involves in the expressions of the macroscopic flow properties (Eq.~\eqref{Macro}) and the spectrum of VDF (Eq.~\eqref{Spectrum}). For numerical analysis, the continuous velocity domain is discretized by $M_\text{vel}=M^1_\text{vel}\times M^2_\text{vel}\times M^3_\text{vel}$ points and the integral is approximated by a certain quadrature rule. The number of velocity grid points is usually larger than the number of frequency components~\cite{wu2014}.

\subsection{Implicit iterative scheme}~\label{Implicit}
In practice, for the steady-state solution of the Boltzmann equation, the derivative of VDF with respect to the time is omitted and the following implicit iterative scheme is usually applied~\cite{WU2013}:
\begin{equation}
\nu^{(t)}f^{(t+1)} + \bm v\cdot\frac{\partial f^{(t+1)}}{\partial\bm x}=\mathcal{C}_{+}^{(t)},
\label{dBoltzmann}
\end{equation}
where the superscripts $(t)$ and $(t+1)$ represent two consecutive iteration steps. The iteration is terminated when the convergence to the steady solution is achieved. For conciseness, we will omit the index of iteration step in the remainder of the paper unless necessary.

\section{The hybridizable discontinuous Galerkin Method}~\label{HDG}

In this section, we present the HDG method for solution of the system ~\eqref{dBoltzmann}. Let $\Delta\in\mathbb{R}^2$ be a two-dimension spatial domain with boundary $\partial\Delta$ in the $x_1-x_2$ plane. $\Delta$ is partitioned into $M_\text{el}$ disjoint regular triangles $\Delta_i$: $\Delta=\cup^{M_\text{el}}_i\Delta_i$. The boundaries $\partial\Delta_i$ of the triangles define a group of $M_\text{fc}$ faces: $\Upsilon=\cup^{M_\text{el}}_i\{\partial\Delta_i\}=\cup^{M_\text{fc}}_c\{\Upsilon_c\}$. The HDG method provides an approximate solution to $f$ on $\Delta_i$ as well as an approximation to its trace $\hat{f}$ on $\Upsilon_c$ in some piecewise finite element spaces $\mathcal{V}\times\mathcal{W}$ of the following forms:
\begin{equation}
\begin{aligned}
\mathcal{V}=\{\varphi_r:\varphi_r|_{\Delta_i}\in\mathcal{P}^k\left(\Delta_i\right),\ r=1,\dots,K_\text{el},\ \forall\Delta_i\subset\Delta\},\\
\mathcal{W}=\{\psi_r:\psi_r|_{\Upsilon_c}\in\mathcal{P}^k\left(\Upsilon_c\right),\ r=1,\dots,K_\text{fc},\ \forall\Upsilon_c\subset\Upsilon\},
\end{aligned}
\end{equation}
where $\mathcal{P}^k\left(D\right)$ denotes the space of $k$-th order polynomials on a domain $D$, $K_\text{el}=\left(k+1\right)\left(k+2\right)/2$ and $K_\text{fc}=k+1$ are the numbers of degree of freedom in triangle and on face, respectively. Then, we have
\begin{equation}
f\left(\bm x,\bm v\right)=\sum^{K_\text{el}}_{r=1}\varphi_rF_r\left(\bm v\right),\quad\hat{f}=\sum^{K_\text{fc}}_{r=1}\psi_r\hat{F}_r\left(\bm v\right),
\label{polynomial}
\end{equation}
where $F_r$ and $\hat{F}_r$ are the degrees of freedom for the VDF and its trace.

\subsection{HDG formulation for the Boltzmann equation}

Introducing $\left(\cdot\right)$ and $\langle\cdot\rangle$ as $\left(a,b\right)_D=\int_{D\subset}\mathbb{R}^2\left(a\cdot b\right)\mathrm{d}x_1\mathrm{d}x_2$ and $\langle a,b\rangle_D=\int_{D\subset\mathbb{R}^1}\left(a\cdot b\right)\mathrm{d}\Upsilon$, respectively, we find the approximation of VDF on $\Delta_i$ such that:
\begin{equation}
-\left(\nabla\varphi_s,\bm vf\right)_{\Delta_i}+\langle\varphi_s,\hat{\bm H}\cdot\bm n\rangle_{\partial\Delta_i}+\left(\varphi_s,\nu f\right)_{\Delta_i}=\left(\varphi_s,\mathcal{C}_{+}\right)_{\Delta_i},\quad\text{for}\ s=1,\dots,K_\text{el},
\label{local}
\end{equation}
where $\bm n$ is the outward unit normal vector, and $\hat{\bm H}$ is the numerical flux defined from the first-order upwind scheme as:
\begin{equation}
\hat{\bm H}\cdot\bm n=\bm v\cdot\bm n\hat{f}+|\bm v\cdot\bm n|\left(f-\hat{f}\right).
\label{flux}
\end{equation}
While, we find the approximation of VDF trace on $\Upsilon_c$ such that the continuity of the normal component of the numerical flux is weakly preserved. On an interior face $\Upsilon_c=\partial\Delta_\text{R}\cap\partial\Delta_\text{L}$ with $\Delta_\text{R}$ and $\Delta_\text{L}$ denoting the right and left triangles at either side of the interface, the continuity is written as:
\begin{equation}
\langle\psi_s,\hat{\bm H}_{\partial\Delta_\text{R}}\cdot\bm n_{\partial\Delta_\text{R}}+\hat{\bm H}_{\partial\Delta_\text{L}}\cdot\bm n_{\partial\Delta_\text{L}}\rangle_{\Upsilon_c}=0,\quad\text{for}\ s=1,\dots,K_\text{fc}.
\label{global}
\end{equation}
Note that, at the boundary face $\Upsilon_c\subset\partial\Delta$, the continuity could be treated in the same way by specifying the flux flowing into the computational domain.

From equations~\eqref{local} and~\eqref{flux}, the solution of $f$ can be expressed as a function of $\hat{f}$, then by eliminating $f$ in the Eq. \eqref{global} and assembling it over all the triangles and faces, we obtain a global matrix system of the form:
\begin{equation}
\mathbb{K}\hat{\mathbf{F}}=\mathbb{R},
\end{equation}
where $\hat{\mathbf{F}}$ is the vector of degrees of freedom of $\hat{f}$.  Once the values of $\hat{f}$ is obtained, the approximation $f$ is recovered from the traces in an element-by-element fashion. The details of the coefficient matrix $\mathbb{K}$ and the right-hand side matrix $\mathbb{R}$, as well as the implementation could be found in the Appendix of~\cite{Su2018}.

\subsection{DG discretization of the collision operator}
Now, we are focusing on the formulation of the terms $\left(\varphi_s,\nu f\right)_{\Delta_i}$ and $\left(\varphi_s,\mathcal{C}_{+}\right)_{\Delta_i}$ in Eq.~\eqref{local}. Inserting the polynomial expansion of $f$ (Eq.~\eqref{polynomial}) into Eq.~\eqref{Spectrum}, the $\bm j$-th spectrum component of the VDF can be rewritten in the polynomial form:
\begin{equation}
\bar{f}^{\bm j}\left(\bm x\right)=\sum^{K_\text{el}}_{r=1}\varphi_r\bar{F}^{\bm j}_r,\quad\bar{F}^{\bm j}_r=\frac{1}{\left(2L\right)^3}\int_{\mathcal{D}}F_r\left(\bm v\right)\exp\left(-\imath\bm\xi^{\bm j}\cdot\bm v\right)\mathrm{d}\bm v,
\end{equation}
where $\bar{F}^{\bm j}_r$ is the spectrum of the degree of freedom of VDF. With some algebraic calculations, the DG discretization of collisional gain term and the collision frequency is expressed as
\begin{equation}
\mathcal{C}_{+}=\sum^{K_\text{el}}_{r=1}\sum^{K_\text{el}}_{p=1}\varphi_r\varphi_p\Xi_{r,p},\quad\nu=\sum^{K_\text{el}}_{r=1}\varphi_r\Lambda_r,
\end{equation}
where,
\begin{equation}
\begin{aligned}
\Xi_{r,p}=\sum_{\bm j=-\bm N/2}^{\bm N/2-1}\sum^{\bm N/2-1}_{\substack{\bm l+\bm m=\bm j\\\bm l,\bm m=-\bm N/2}}\bar{F}^{\bm l}_r\bar{F}^{\bm m}_p\beta\left(\bm l,\bm m\right)\exp\left(\imath\bm\xi^{\bm j}\cdot\bm v\right),\\
\Lambda_r=\sum_{\bm j=-\bm N/2}^{\bm N/2-1}\bar{F}^{\bm j}_r\beta\left(\bm j,\bm j\right)\exp\left(\imath\bm\xi^{\bm j}\cdot\bm v\right).
\end{aligned}
\end{equation}

Finally, we obtain that
\begin{equation}
\left(\varphi_s,\nu f\right)_{\Delta_i}=\sum^{K_\text{el}}_{r=1}\sum^{K_\text{el}}_{p=1}\Lambda_rF_p\left(\varphi_s,\varphi_r\varphi_p\right)_{\Delta_i},
\label{Direct}
\end{equation}
\begin{equation}
\left(\varphi_s,\mathcal{C}_{+}\right)_{\Delta_i}=\sum^{K_\text{el}}_{r=1}\sum^{K_\text{el}}_{p=1}\Xi_{r,p}\left(\varphi_s,\varphi_r\varphi_p\right)_{\Delta_i},
\end{equation}

\subsection{Reduction of computation in collision term}
By applying the FFT-based convolution, the computational cost of $\Xi_{r,p}$ is $O\left(M^2_\text{qua}N\log\left(N\right)\right)$. Therefore, the total cost to evaluate the collisional gain term $\mathcal{C}_{+}$ on one triangle $\Delta_i$ is equal to $O\left(K^2_{\text{el}}M^2_\text{qua}N\log\left(N\right)\right)$, which could dramatically increase when high-order discretization is applied. Actually, the computational cost could be reduced in the following way. We choose the basis function as nodal shape functions:
\begin{equation}
\varphi_r\left(\bm x_p\right)=\begin{cases}
0,\quad\text{if}\ r\neq p,\\
1,\quad\text{if}\ r=p,
\end{cases}
\end{equation}
where, $\bm x_p$ is the nodal points for interpolation, thus $F_r=f\left(\bm x_r\right)$ are the nodal values of VDF. It is assumed that the distribution of $\mathcal{C}_{+}$ within a triangle might as well be estimated by the nodal approximation:
\begin{equation}
\mathcal{C}_{+}\simeq\mathcal{\tilde{C}}_{+}=\sum^{K_\text{el}}_{r=1}\varphi_r\tilde{\Xi}_r,
\label{Reduction}
\end{equation}
where its nodal values are related to the corresponding nodal values of VDF as $\tilde{\Xi}_r=\Xi_{r,r}$. As a consequence the calculation cost of $\left(\varphi_s,\mathcal{\tilde{C}}_{+}\right)_{\Delta_i}=\sum^{K_\text{el}}_{r=1}\tilde{\Xi}_r\left(\varphi_s,\varphi_r\right)_{\Delta_i}$ is reduced to $O\left(K_{\text{el}}M^2_\text{qua}N\log\left(N\right)\right)$.


\section{Results and Discussions}\label{Results}

For verification, the HDG method of $k$ up to 4 is applied to solve the linearized BGK equation. The convergence criterion for the iterative procedure described in Sec.~\ref{Implicit} is that the global relative residual in flow property $\mathcal{Q}$ between two successive iteration steps is less than a threshold value $\epsilon$. The residual is defined as
\begin{equation}
R_{\mathcal{Q}}=\frac{|\int\mathcal{Q}^{\left(t+1\right)}-\mathcal{Q}^{\left(t\right)}\mathrm{d}x_1\mathrm{d}x_2|}{|\int\mathcal{Q}^{\left(t\right)}\mathrm{d}x_1\mathrm{d}x_2|}
\end{equation}
In the following cases, the convergence tests in terms of the discrete velocities are performed first to determine the number of points in the molecular velocity space and frequency domain: the convergence is said to be reached if further refinement of the grid would only improve the solutiosn by a magnitude no more than 0.5\%. The entire tests are dome in double precision on a workstation with Intel Xeon-E5-2680 processors and 132 GB RAM. During iteration, we call the relative routines in Intel Math Kernel Library (MKL) to invert the matrix. More over to solve the HDG global equations, we call the iterative sparse solver, Intel MKL PARDISO, which is based on the Conjugate gradients squared method. The criterion for solving the linear system is set as $10^{-3}$. The first tests are done on single processor, and the int internal parallelism for MKL functions are also not activated. The second simulations are rum on multiple processors using OpenMP. The results from solver with direct calculation Eq.~\eqref{Direct} are labeled as `HDG-d', while the ones from solver with reduction scheme Eq.~\eqref{Reduction} are labeled as `HDG-r'.

\subsection{Planar Couette flow}

\begin{figure}[t]
	\begin{centering}
		\includegraphics[width=0.96\textwidth]{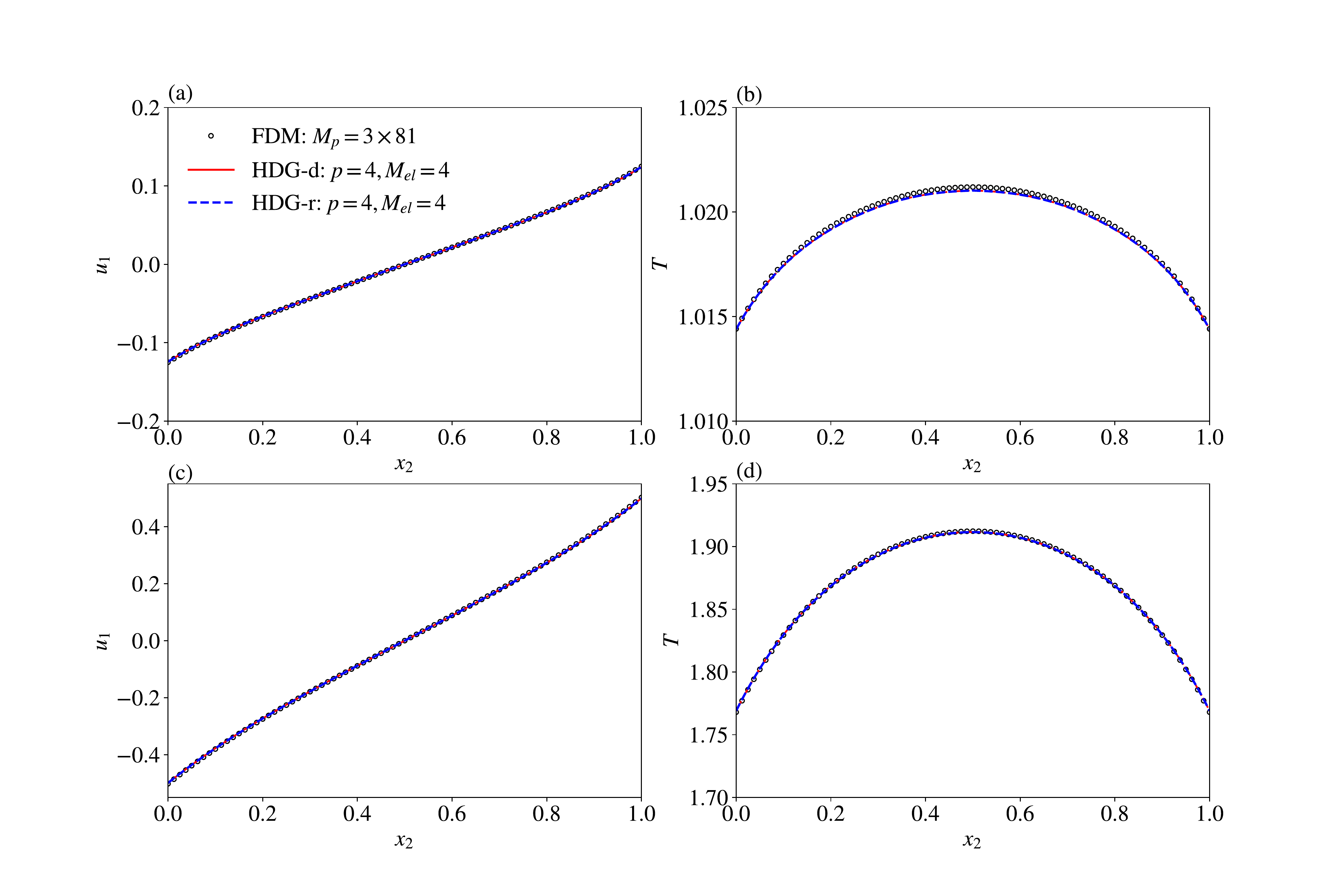}
		\par\end{centering}
	\caption{Profiles of velocity and temperature for planar Couette flow of argon gas at: (a) and (b) $Kn=0.5$, $U_\text{w}=0.2$; (c) and (d) $Kn=2.0$, $U_\text{w}=1.2$.}
	\label{Couette}
\end{figure}

Couette flow between two parallel plates with a distance of $H$ is used to assess accuracy and efficiency of the proposed HDG solver. The one-dimensional flow is resolved on a two-dimensional (2D) domain with a column of uniform isosceles right triangles being set along the direction perpendicular to the plates, say, the $x_2$ direction. The wall temperature is set as the reference temperature as $T_0=273$ K. The flow gas is argon with a shear viscosity proportional to $T^{0.81}$. We consider two cases, one is at $Kn=0.5$ with a wall velocity of $U_\text{w}=0.2$, while the other is at $Kn=2.0$ with $U_\text{w}=1.2$. The truncated molecular domain is $[-6,6]^3$, and $32\times32\times24$ ($52\times52\times24$) velocity points are used for case of $Kn=0.5$ ($Kn=2.0$). The HDG solutions are compared with ones from a second-order finite difference method (FDM), which has been verified by the direct simulation Monte Carlo (DSMC) results for this canonical problem~\cite{WU2013}.

The velocity and temperature profiles from different solvers are shown in Figure~\ref{Couette}, in which $M_\text{p}$ denotes the number of equidistant points used in the spatial space for the FDM. The velocity grid for the FDM is the same mentioned above. For the 2nd-order scheme, at least 3 points are required in the $x_1$ direction on the 2D computational domain. It is found that the HDG solver with reduced calculation of collision operator produces the same results as those of the one with full calculation of collision operator. The HDG results agree very well with the FDM, where maximum discrepancy of $0.016\%$ appears in $T$ in the smaller Knudsen number case.

To compare the performance of the HDG-d and HDG-r, we list the half-channel mass flow rate (MFR) $Q=\int^{0.5}_0 u_1\mathrm{d}x_2$, the number of iterative step to obtain the converged solution and the CPU time $t_\text{c}$ for both the schemes with different order of approximation polynomials in Table~\ref{CouetteHDG}. Actually, the HDG-r uses the same number of iterative steps to obtain the results, so we don't list it in the table. Both schemes give nearly the same half-channel MFR. However, the CPU time for HDG-r is significantly less than that of the HDG-d, especially for higher-order approximation. For $k=4$, the HDG-r is more than 6 times faster than the HDG-d. We also list the results for the FDM in Table~\ref{CouetteFDM}. It is found that the HDG-r scheme is more efficient than the FDM. For example, for case of $Kn=0.5$, the HDG method obtains a converged MFR of about 2.844, while the FDM obtain a value of about 2.847. To reach such a converged solutions, the HDG-r with $k=4$ uses a spatial grid with only 4 triangles and costs 96.9 seconds, while the FDM requires $3\times401$ points for the spatial discretization and 595.4 seconds.

\begin{sidewaystable}
\caption{Comparisons between the HDG-d (with direct calculation of $\mathcal{C}_{+}$) and HDG-r (with reduced calculation of $\mathcal{C}_{+}$) in terms of the half-channel mass flow rate ($Q$), the number of iterations (Itr denotes the number of iteration steps to reach the convergence criterion $R_{u_1}<10^{-5}$), and the CPU time $t_\text{c}$. Couette flow between two parallel plates is considered. }

\centering{}%
\begin{tabular}{ccccccccccccccc}
\hline
\multicolumn{8}{c}{$Kn=0.5,\ U_{\text{w}}=0.2$} &  & \multicolumn{6}{c}{$Kn=2.0,\ U_{\text{w}}=1.2$}\tabularnewline
\hline
\multicolumn{5}{c}{HDG-d} &  & \multicolumn{2}{c}{HDG-r} &  & \multicolumn{3}{c}{HDG-d} &  & \multicolumn{2}{c}{HDG-r}\tabularnewline
\cline{1-5} \cline{7-8} \cline{10-12} \cline{14-15}
$k$ & $M_{\text{el}}$ & $Q\times10^{-2}$ & Itr & $t_{c}$, {[}s{]} &  & $Q\times10^{-2}$ & $t_{c}$, {[}s{]} &  & $Q\times10^{-1}$ & Itr & $t_{c}$, {[}s{]} &  & $Q\times10^{-1}$ & $t_{c}$, {[}s{]}\tabularnewline
\hline
\multirow{3}{*}{1} & 4 & 2.9009 & 32 & 32..5 &  & 2.9009 & 14.6 &  & 1.1861 & 25 & 106.4 &  & 1.1816 & 41.1\tabularnewline
 & 8 & 2.8520 & 30 & 62.6 &  & 2.8521 & 30.4 &  & 1.1664 & 59 & 499.3 &  & 1.1653 & 191.5\tabularnewline
 & 16 & 2.8462 & 50 & 223.9 &  & 2.8462 & 109.6 &  & 1.1640 & 113 & 1767.8 &  & 1.1637 & 775.1\tabularnewline
\hline
\multirow{3}{*}{2} & 4 & 2.8399 & 31 & 114.3 &  & 2.8399 & 29.8 &  & 1.1618 & 31 & 439.9 &  & 1.1619 & 102.7\tabularnewline
 & 8 & 2.8445 & 30 & 227.8 &  & 2.8445 & 64.1 &  & 1.1634 & 61 & 1813.1 &  & 1.1633 & 410.4\tabularnewline
 & 16 & 2.8446 & 50 & 829.4 &  & 2.8446 & 223.0 &  & 1.1635 & 114 & 6152.0 &  & 1.1635 & 1587.2\tabularnewline
\hline
\multirow{3}{*}{3} & 2 & 2.8369 & 25 & 118.4 &  & 2.8366 & 22.2 &  & 1.1621 & 16 & 320.4 &  & 1.1614 & 46.2\tabularnewline
 & 4 & 2.8455 & 31 & 301.3 &  & 2.8455 & 55.5 &  & 1.1634 & 31 & 1233.5 &  & 1.1633 & 183.9\tabularnewline
 & 8 & 2.8444 & 30 & 651.6 &  & 2.8444 & 117.3 &  & 1.1634 & 61 & 4332.5 &  & 1.1634 & 728.6\tabularnewline
\hline
\multirow{3}{*}{4} & 2 & 2.8394 & 25 & 246.7 &  & 2.8395 & 38.0 &  & 1.1620 & 25 & 1031.8 &  & 1.1621 & 118.7\tabularnewline
 & 4 & 2.8444 & 31 & 696.2 &  & 2.8444 & 96.9 &  & 1.1633 & 31 & 2631.5 &  & 1.1633 & 309.5\tabularnewline
 & 8 & 2.8444 & 30 & 1344.6 &  & 2.8444 & 197.2 &  & 1.1634 & 61 & 9176.8 &  & 1.1634 & 1212.1\tabularnewline
\hline
\end{tabular}
\label{CouetteHDG}
\end{sidewaystable}

\begin{table}[t]
\caption{Couette flow between two parallel plates solved by the FDM. $M_\text{p}$ is the number of discrete points in the spatial space, $Q$ is the half-channel mass flow rate, Itr is the number of iteration steps to satisfy the convergence criterion $R_{u_1}<10^{-5}$, and $t_\text{c}$ is the CPU time.}

\begin{centering}
\begin{tabular}{cccccccc}
\hline
\multicolumn{4}{c}{$Kn=0.5,\ U_{\text{w}}=0.2$} &  & \multicolumn{3}{c}{$Kn=2.0,\ U_{\text{w}}=1.2$}\tabularnewline
\cline{1-4} \cline{6-8}
$M_{\text{p}}$ & $Q\times10^{-2}$ & Itr & $t_{c}$, {[}s{]} &  & $Q\times10^{-2}$ & Itr & $t_{c}$, {[}s{]}\tabularnewline
\hline
$3\times21$ & 2.9045 & 66 & 26.1 &  & 1.1952 & 23 & 36.8\tabularnewline
$3\times41$ & 2.8715 & 64 & 54.4 &  & 1.1776 & 22 & 90.3\tabularnewline
$3\times81$ & 2.8570 & 63 & 112.7 &  & 1.1670 & 22 & 161.6\tabularnewline
$3\times161$ & 2.8505 & 63 & 227.5 &  & 1.1665 & 22 & 321.4\tabularnewline
$3\times201$ & 2.8493 & 63 & 310.8 &  & 1.1659 & 22 & 388.8\tabularnewline
$3\times321$ & 2.8475 & 63 & 483.1 &  & 1.1649 & 22 & 588.4\tabularnewline
$3\times401$ & 2.8469 & 63 & 595.4 &  & 1.1646 & 22 & 728.0\tabularnewline
$3\times501$ & 2.8465 & 63 & 741.2 &  & 1.1643 & 22 & 892.4\tabularnewline
\hline
\end{tabular}
\par\end{centering}
\label{CouetteFDM}
\end{table}

\subsection{Lid-driven cavity flow}

\begin{figure}[t]
	\begin{centering}
		\includegraphics[width=0.96\textwidth]{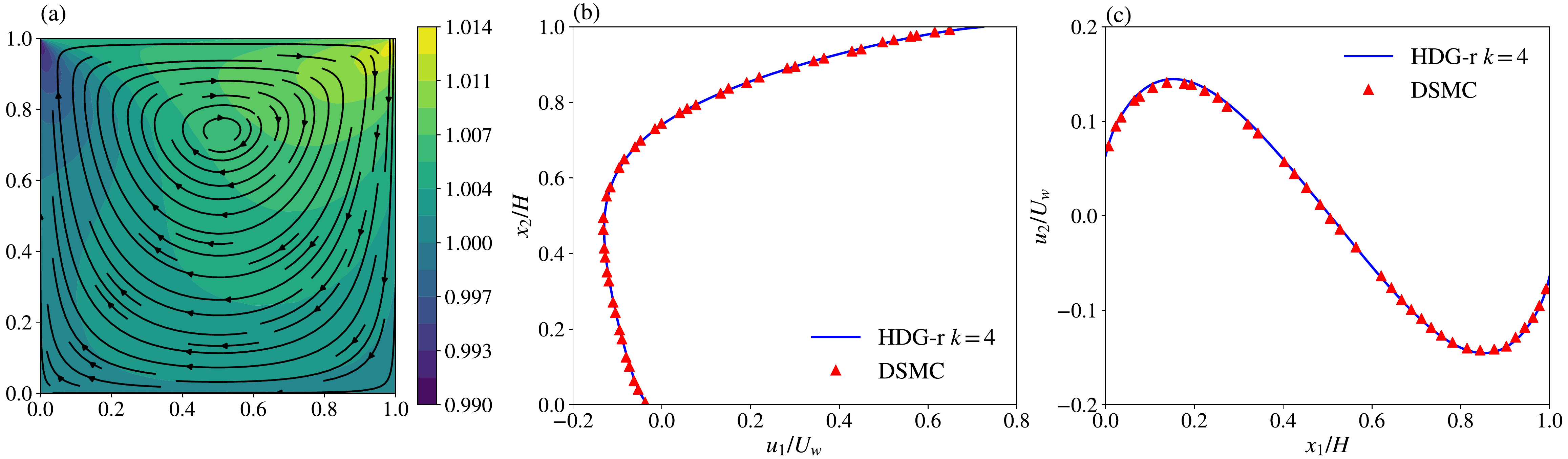}
		\par\end{centering}
	\caption{Lid-driven cavity flow at $Kn=0.1$: (a) temperature contour and stream lines; (b) horizontal velocity along the vertical central line; (c) vertical velocity along the horizontal central line.}
	\label{Cavity}
\end{figure}

By comparing with the DSMC results, a 2D gaseous flow driven in a square cavity driven by the top lid is used to further assess accuracy of the HDG scheme of reduced calculation of the collision operator. The cavity has a dimension of $H\times H$. The wall temperature is set as the reference temperature as $T_0=273$ K. The velocity of the driven lid is 0.148 in dimensional form (or 50 m/s). The flow gas is argon with a shear viscosity index of 0.81. The gas flow is initialed to be rest at $T_0$ with $Kn=0.1$. The truncated molecular domain is $[-6,6]^3$ and $32\times32\times24$ discrete velocities are employed. For the spatial discretization, total 392 triangles are used. The closer to the driven lid, the smaller the triangle size. It takes about 174 iterative steps to approach to the steady-state solution with $\max\left(R_{u_1},R_{u_2}\right)<10^{-5}$. Figure~\ref{Cavity} shows the temperature contour, stream lines and horizontal (vertical) velocity along the vertical (horizontal) central line. The velocity profiles are compared with the DSMC results~\cite{John2010}. The HDG-r results agree well with the DSMC ones.

\section{Conclusions}\label{Concludsion}

In summary, we have applied the high-order hybridizable discontinuous Galerkin discretization to solve the Boltzmann equation with full collision integral. An implicit iterative scheme is employed to find the steady-state solutions. The molecular velocity distribution function and its trace are approximated on arbitrary triangular spatial mesh and the mesh skeleton, respectively. By imposing the continuity of the normal flux on the triangle faces, a final global systems for VDF traces are obtained with fewer coupled degree of freedom compared to the classical DG method. The fast spectral method is used to evaluate the collision operator with general intermolecular potentials. The DG discretization is incorporated into the fast spectral method. By introducing a special polynomial approximation to the collision operator, the computational cost for the collision operator within a triangle is proportional to $O\left(K_\text{el}M^2_\text{qua}N\log N\right)$. Two different validation problem have been presented to show accuracy and capability of the prosed scheme. By comparing with the FDM and DSMC results, it is demonstrated that the HDG scheme is accurate and more efficient than the FDM.

\section*{Acknowledgments}

This work is jointly founded by the Royal Society of Edinburgh and National Natural Science Foundation of China under Grant No. 51711530130. It is also financially supported by the Carnegie Research Incentive Grant for the Universities in Scotland, and the Engineering and Physical Sciences Research Council (EPSRC) in the UK under grant EP/M021475/1.

\section*{References}

\bibliographystyle{elsarticle-num}
\bibliography{mybibfile}

\end{document}